\documentclass{npw}
\usepackage{graphicx,color}
\usepackage{amsmath}

\def\be{\begin{equation}}
\def\ee{\end{equation}}
\def\bea{\begin{eqnarray}}
\def\eea{\end{eqnarray}}
\def\rmd{{\rm d}}
\def\Re{{\rm Re}}

\def\bfr{{\bf r}}
\def\bfp{{\bf p}}

\def\abar{{\tilde a}}
\def\abar3{{\tilde a}_3}
\def\abar5{{\tilde a}_5}
\def\abar6{{\tilde a}_6}

\def\exp{\hbox{exp}}

\def\vareps{E}

\def\ramaAM{\vadjust{\vbox to 0pt{\vss \hbox to \hsize
{\hskip\hsize \quad $\Leftarrow$\quad {\it AM}\hss}\vskip3.5pt}}}
\def\ramaKA{\vadjust{\vbox to 0pt{\vss \hbox to \hsize
{\hskip\hsize \quad $\Leftarrow$\quad {\it KA}\hss}\vskip3.5pt}}}
\def\ramaNK{\vadjust{\vbox to 0pt{\vss \hbox to \hsize
{\hskip\hsize \quad $\Leftarrow$\quad {\it NK}\hss}\vskip3.5pt}}}
\def\ramaYK{\vadjust{\vbox to 0pt{\vss \hbox to \hsize
{\hskip\hsize \quad $\Leftarrow$\quad {\it NK}\hss}\vskip3.5pt}}}
\def\ramaMB{\vadjust{\vbox to 0pt{\vss \hbox to \hsize
{\hskip\hsize \quad $\Leftarrow$\quad {\it MB}\hss}\vskip3.5pt}}}

\def\siml{\hbox{\kern.1em \lower.6ex \hbox{$\sim$} \kern-1.12em
          \raise.6ex \hbox{$<$} \kern.1em }}
\def\simg{\hbox{\kern.1em \lower.6ex \hbox{$\sim$} \kern-1.12em
          \raise.6ex \hbox{$>$} \kern.1em }}
\begin{document}
\title{Semiclassical treatment of symmetry breaking\\ 
and bifurcations in a 
non-integrable potential}
\author{
M V Koliesnik, Ya D Krivenko-Emetov, A G Magner\email{magner@kinr.kiev.ua}\\
{\it Institute for Nuclear Research,  03680 Kyiv, Ukraine}\\
and {\it National Technical University of Ukraine, 03056 Kyiv}\\
K Arita\\
{\it Department of Physics, Nagoya Institute of Technology,
         Nagoya 466-8555, Japan}\\
M Brack\\
{\it  Institute for Theoretical Physics, University of Regensburg, 
 D-93040 Regensburg, Germany }\\
}
\pacs{03.65.Ge, 03.65.Sq, 05.45.Mt, 21.60.Cs}

\date{}
\maketitle

\begin{abstract}

We have derived an analytical trace formula for the level density 
of the H\'enon-Heiles 
potential using the improved stationary phase method, based on 
extensions of Gutzwiller's semiclassical path integral approach.
This trace formula has the correct limit to the standard 
Gutzwiller trace formula for the isolated periodic orbits 
far from all (critical) symmetry-breaking points.
It continuously joins all critical points at which an 
enhancement of the semiclassical amplitudes occurs. 
We found a good agreement between the semiclassical and the quantum 
oscillating level densities 
for the gross shell structures and for the energy shell
corrections, solving the symmetry breaking problem at small energies. 

{\bf Keywords:} Periodic orbit theory,
Gutzwiller trace formula, symmetry breaking, level density and energy
shell corrections.

\end{abstract}

\section{Introduction}

Semiclassical periodic orbit theory (POT) is a convenient tool for analytical 
studies of the shell structure in the single-particle level density of finite
fermionic systems \cite{gutz,strut,strutmag,strusem,book,migdal}. This theory
relates the level density and energy shell corrections to the sum of 
periodic orbits and their stability characteristics, and thus, gives 
the analytical quantum-classical correspondence.  
Recent studies of the POT focused on overcoming catastrophe problems in the 
derivation of the semiclassical trace formulae arising in connection with
symmetry breaking and bifurcation phenomena, where the standard stationary
phase method (SPM) fails (see \cite{gutz}). Semi-analytical 
uniform approximations 
solving these problems for the case of well separated pichfork bifurcations
in the non-integrable H\'enon-Heiles (HH) potential \cite{hh} were suggested 
in \cite{kaidelbrack,bracktanaka}, using the normal-form 
theory of non-linear dynamics \cite{sie97,ssun97,ssun98,schomerus}.

\begin{figure*}[t] 
\begin{center}
\includegraphics[width=0.8\textwidth,clip]{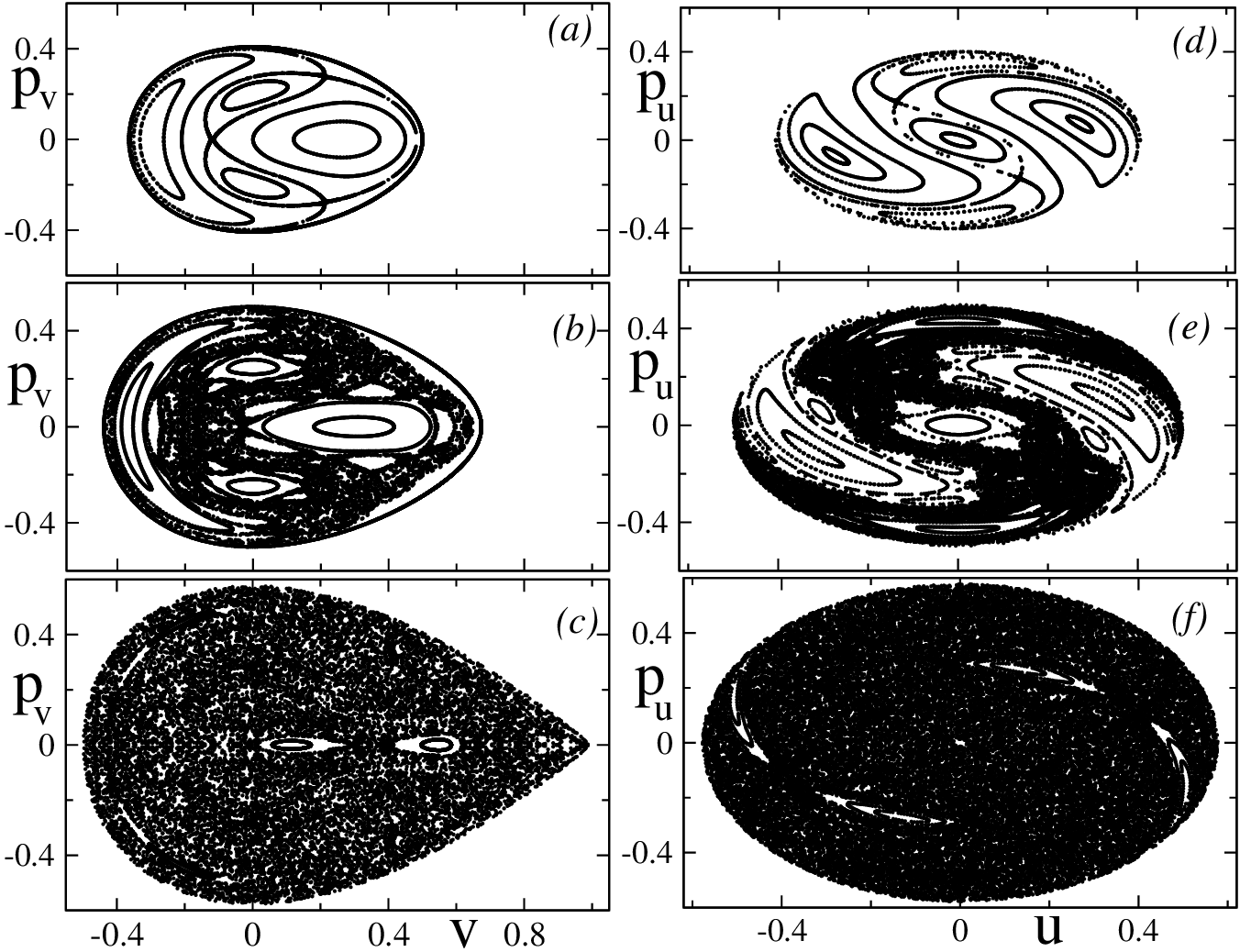}
\end{center}
\caption{Poincar\'e surfaces of sections (PSS) of the 
scaled H\'enon-Heiles Hamiltonian $h$ (\ref{scaling});
left column: {\it (a), (b)} and {\it (c)} plots show the PSS
at $u=0$ for the energies 
$e=0.5$, $0.75$ and $1.0$, respectively; right column: 
{\it (d), (e)} and {\it (f)} graphics are given for $v=0$
at the same energies. 
 }
\label{fig1}
\end{figure*}

In the present work, we derive an analytical semiclassical trace formula 
for the level density of the HH potential, employing the improved 
stationary phase method (ISPM) \cite{mfaptp2006} valid for arbitrarily dense 
sequences of pitchfork bifurcations near the saddle-point energy and for 
harmonic-oscillator symmetry breaking in the limit of small energies. 
In this respect, the regular-to-chaotic transition in Fermi systems
becomes important for the understanding of its influence on shell
correction amplitudes. Fig.\ \ref{fig1} shows transparently
such a transition through Poincar\'e Surfaces of Section (PSS) 
of the non-linear classical dynamics for the 
HH potential as a simple non-trivial example 
\cite{book,hhprl}, 
see also
\cite{BMYnpae2010,BMYijmp2011,BMprc2012} for the PSS and Lyapunov 
exponents in the three-dimensional 
axially symmetric Legendre-polynomial and spheroidal billiards. 
As shown in this figure, the obvious transition from chaos to order occurs
with decreasing energy $e$ of the particle from the saddle ($e=1$) to 
a small energy (harmonic oscillator) limit $e \rightarrow 0$.
We show below the relation of this behavior of the PSS to the amplitudes of 
oscillations in the level density (density of states) and total energy of
fermion systems. 

\section{Periodic-orbit theory in phase space}

The level density $g(E)$ is obtained from the semiclassical
Green's function \cite{gutz} by taking its trace
 in the phase-space Poincar\'e variables $Q,p$  
\cite{sie97,ssun97,ssun98,schomerus,mfaptp2006}:
\begin{align}\label{pstraceqP}
g_{\rm sc}(E)=& \frac{1}{(2\pi\hbar)^2}\Re\sum_{\rm ct}
\int\! {\rm d}Q\int\! {\rm d}p\; T_{y\,{\rm ct}} \,
\left|\mathcal{J}_{\rm ct}(p,P)\right|^{1/2}\nonumber\\
 & \times\,\exp\left\{\frac{i}{\hbar}
\left[\widehat{S}_{ct}(Q,p,\vareps) - Q p\right] - 
{i \pi \over 2} \mu_{\rm ct}\right\}\!.
\end{align}
Here $Q$ and $p$ are the final $x''$ and initial $p_x'$ coordinates 
in the phase-space variables 
$x,y,p_x,p_y$ perpendicular to a reference classical
trajectory (ct) in two dimensions,
  $T_{y\,{\rm ct}}=m\oint d y/p_y$ is the primitive partial period of 
the $y$ motion along the ct, $\widehat{S}_{\rm ct}(Q,p,\vareps)$ the generating
function, $\mu_{\rm ct}$ the Maslov phase, and 
$\mathcal{J}_{\rm ct}(p,P)$ is the Jacobian for the transformation between the 
variables shown as its arguments. The ISPM generating function  
$\widehat{S}_{ct}(Q,p,\vareps)$ is defined by
\be\label{genfunact}
{\widehat S}_{\rm ct}(Q,p,\vareps) = S_{\rm ct}(Q,p,\vareps)  + qp ,
\ee
where  $~S_{\rm ct}(Q,p,\vareps)~$ is the action 
$~~~S_{\rm ct}(\bfr',\bfr'',\vareps)=$\\
$=\int_{\bfr'}^{\bfr''} \bfp\cdot\rmd \bfr$ 
expressed in terms of
the Poincar\'e variables $Q$ and $p$ through the mapping transformation
equations $Q=Q(q,p)$ and $P=P(q,p)$ along a ct ($\bfr'$ and $\bfr''$ are the
initial and final spatial coordinates of the ct).
It can be replaced by a (truncated) fourth-order expansion 
around the stationary points $Q^*, p^*$  which correspond to the periodic 
orbits (POs), $Q^*=q~$, $p^*=P$  \cite{mfaptp2006}. For 
pichfork bifurcations, the expansion of the generating function
${\widehat S}_{\rm ct}(Q,p,\vareps)$ (\ref{genfunact}) is  
similar to the normal forms \cite{sie97,ssun97,ssun98,schomerus} with 
the following power series in $Q-Q^*$ and $p-p^*$:
\begin{align}\label{genfunexp}
& S_{\rm ct}(Q,p,\vareps) =
S_{\rm PO}(\vareps) + 
\epsilon_{{}_{\! {\rm PO}}}^{(Q)}\, (Q-Q^*)^2  
\nonumber\\
& \quad +a_{{}_{\! {\rm PO}}}^{(Q)}\,(Q-Q^*)^4 + 
 \epsilon_{{}_{\! {\rm PO}}}^{(p)} (p-p^*)^2
 + a_{{}_{\! {\rm PO}}}^{(p)} (p-p^*)^4,\;
\end{align}
where $S_{\rm PO}(\vareps)$ is the action along the PO.
 Performing also more exact integrations 
over $Q$ and $p$ in (\ref{pstraceqP}), one obtains
for the case of pitchfork bifurcations 
\begin{align}\label{dgscl}
\delta g_{\rm sc}(\vareps)=&\frac{1}{(2 \pi \hbar)^2}\; 
\Re\sum_{\rm PO} \frac{T_{\rm PPO}}{[\hbar^2 a_{\rm PO}^{(Q)}a_{\rm PO}^{(p)}]^{1/4}}\; 
\mathcal{A}\left(\xi_{\rm PO}^{(Q)}\right)\nonumber\\
&\times\, \mathcal{A}\!\left(\xi_{\rm PO}^{(p)}\right)\!
\exp\!\left[\frac{i}{\hbar}S_{\rm PO}(\vareps) -\frac{i\pi}{2} 
\sigma_{{}_{\! {\rm PO}}}\! 
-i \phi\right]\!,\;
\end{align}
where $T_{\rm PPO}$ is the period for a primitive PO, $S_{PO}(E)$ is 
its full action
(including repetitions) at energy $E$, and 
\be\label{amphh}
\mathcal{A}(\xi)=\int_{z_{-}}^{z_{+}} \rmd z\;
\exp\left[i \left(\xi z^2 + z^4\right)\right], \quad \xi=
\frac{\varepsilon}{(\hbar a)^{1/2}},
\ee
is the amplitude factor; $\varepsilon$ and $a$ (for $a>0$) are 
the coefficients in the power expansion  
of the generating function $\widehat{S}_{ct}(Q,p,\vareps)$ 
(\ref{genfunact}) with (\ref{genfunexp}) 
in $Q-Q^*$ and $p-p^*$, which are proportional to the 2nd and 4th  
derivatives of 
$\widehat{S}_{ct}(Q,p,\vareps)$) at the stationary points $Q^*$ and $p^*$;
$\sigma_{\rm PO}$ is the Maslov index related to the turning and caustic points 
along the POs, $\phi$ a constant phase independent of the PO. The integration 
(\ref{amphh}) is performed over the finite classically accessible region of the
Poincar\'e variables $Q$ and $p$, denoted here as 
\be\label{zQpeq}
z=\frac{Q-Q^*}{(a^{(Q)}/\hbar)^{1/4}},\qquad
z=\frac{p-p^*}{(a^{(p)}/\hbar)^{1/4}},
\ee
i.e.\ from $z_{-}^{(Q)}$ to $z_{+}^{(Q)}$ and from $z_{-}^{(p)}$ to $z_{+}^{(p)}$, 
respectively, with
\be\label{zpmQp}
z_{\pm}^{(Q)}=\frac{Q_\pm-Q^*}{(a^{(Q)}/\hbar)^{1/4}},\qquad
z_{\pm}^{(p)}=\frac{p_\pm-p^*}{(a^{(Q)}/\hbar)^{1/4}}.
\ee
In (\ref{dgscl}), the sum runs over the straight-line orbits $A_\sigma$, the 
rotational orbits $R_\sigma$, and the librational orbits $L_\sigma$ of the standard 
HH Hamiltonian \cite{hh} (here in units with $m=\omega=\hbar=1$):
\be\label{hhpoten}
H=\frac12\left(\dot{x}^2+\dot{y}^2\right) + 
\frac12\left(x^2+y^2\right) + \alpha\left(x^2 y -
\frac{y^3}{3}\right).
\ee
\begin{figure*}[t]
\begin{center}
\includegraphics[width=0.8\textwidth,bb=30 40 760 380,clip]{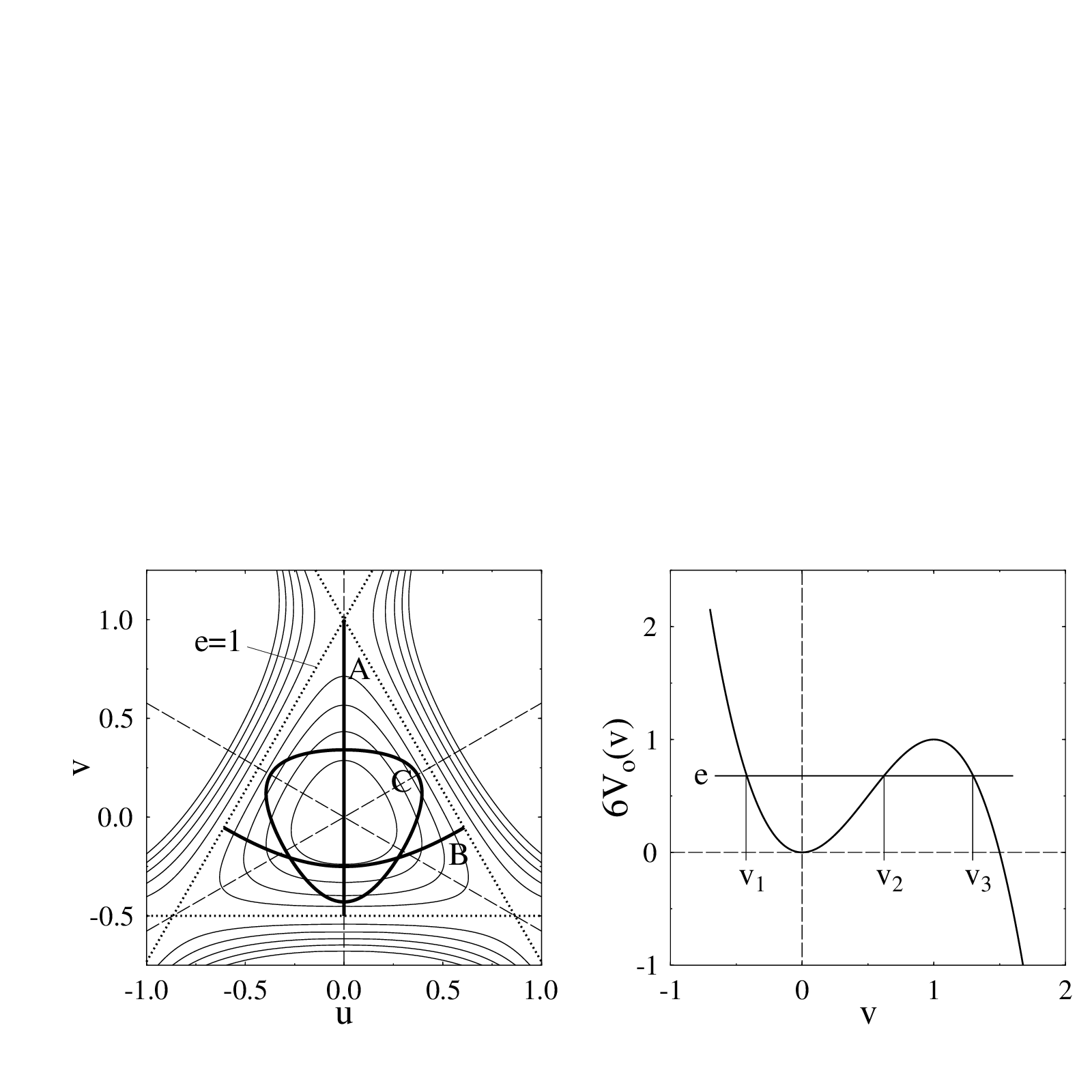}
\end{center}
\caption{The scaled H\'enon-Heiles potential 
of the Hamiltonian (\ref{scaling}). 
{\it Left:} Equipotential contour lines are given in scaled energies $e$ in the 
plane $(u,v)$. The dashed lines are the symmetry axes. The three shortest orbits 
A, B, and C (evaluated at the energy $e=1$) are shown by the heavy solid lines. 
{\it Right:} Cut along $u=0$ shows a barrier. (After \cite{hhprl}.) 
}
\label{fig2}
\end{figure*}

Using the barrier energy $E_{barr}=1/(6\alpha^2)$ as dimensionless energy unit,
$e=E/E_{barr}=6 \alpha^2 E$, and the following scaled variables
\bea\label{scaling}
p_u&\!\!=\!\!&\alpha p_x\,,\quad p_v=\alpha p_y\,,\quad u=\alpha x\,, 
\quad v=\alpha y\,,\nonumber\\
{}\hspace{-2.0ex}h&\!\!=\!\!&3(\dot{u}^2+\dot{v}^2)+ 3(u^2+v^2)+6vu^2-2v^3,
\eea
one obtains classical dynamic equations independent of the parameter $\alpha$
\be\label{eqmot}
\ddot{u}=-u-2 u v\,,\qquad \ddot{v}=-v + v^2 - u^2.
\ee

The scaled HH potential is shown in Fig.\ \ref{fig2}
as equipotential lines, and the orbits A, B and C (at $e=1$) are presented, too.
The HH potential is invariant under rotations about 120 degrees, which leads to
a discrete degeneracy of the orbits. Such a degeneracy can be
simply taken into account multiplying the amplitudes in the trace formula
(\ref{dgscl}) by a factor 3. The cut along $u=0$ (right) shows a barrier 
at the saddle $e=1$ with two turning points $v_1 \leq v_2$ at $0<e< 1$;
$v_n$ are the real solutions of the cubic equation $e=3 v^2- 2 v^3\leq 1$
(\ref{yn}).

In order to simplify the amplitude function $\mathcal{A}$ in the ISPM 
trace formula (\ref{dgscl}), we note that sufficiently far from the 
symmetry-breaking limit $E=0$, the integration limits in (\ref{amphh})
can be extended to $\pm \infty$ (convergence being guaranteed by the
finite fourth-order terms). Then, the amplitudes $\mathcal{A}$ (\ref{amphh})
can be expressed through integral representations of the Bessel functions
$J_{\pm 1/4}(x)$: 
\begin{align}\label{amphhpp}
\mathcal{A}(\xi)=&\;
\frac{\pi}{2}\; \sqrt{\xi}\;
\left\{\exp\left[-i \left(\frac{\xi^2}{8}-\frac{\pi}{8}\right)\right]
J_{-1/4}\left(\frac{\xi^2}{8}\right)\right.
\nonumber\\
&-\; \frac{\xi}{|\xi|}\,\left.\exp\left[-i \left(\xi +
\frac{\pi}{8}\right)\right]
J_{1/4}\left(\frac{\xi^2}{8}\right) \right\}.
\end{align}
Here we took into account
a time-reversal symmetry by inclusion of the factor 2 where necessary. 
Note that more exact trace formulae (with additional terms
proportional to Bessel functions with indices $\pm\, 3/4$ etc.) 
can be derived by taking into account higher-order terms in the phase
and amplitude factors, respectively. This gives results similar to those
obtained in \cite{ssun97,ssun98} using the normal forms for pitchfork
bifurcations. 

Using asymptotic forms of the Bessel functions for large arguments $\xi$
in (\ref{amphhpp}), one obtains from (\ref{dgscl}) (with $\phi=0$) the 
standard Gutzwiller trace formula 
\cite{gutz,book} valid for isolated POs:
\begin{gather}
\delta g_{\rm sc}(E)\rightarrow 
\sum_{\rm PO} \mathcal{A}_{\rm PO}^{G}(E) 
\cos\left[\frac{S_{\rm PO}(E)}{\hbar} \!-\!\frac{i\pi}{2} 
\sigma_{{}_{\! {\rm PO}}}\right],\nonumber\\
\mathcal{A}_{\rm PO}^{G}(E)=T_{\rm PPO}\left(\pi \hbar
\sqrt{\Big|2 - {\rm Tr\,\mathcal{M}}_{\rm PO}\Big|} \right)^{\!-1},
\label{gutz}
\end{gather}
where $\mathcal{M}_{\rm PO}$ is the stability matrix for the PO.
Numerical and
analytical calculations and the remarkable 
``fan'' structure of the pitchfork bifurcations of the straight-line
orbits $A_\sigma$ were analyzed in the case of the HH potential
\cite{bmtjpa2001,bkwf2006,fmbpre2008}. As shown in the Appendix 
(Fig.\ \ref{fig6}), 
several analytical approximations for $TrM_A$ can be derived in terms of the
simplest Mathieu functions for smaller energies $e \siml 0.8$,
and in terms of improved Legendre solutions for the whole region
from the zero energy to the saddle, $0 \leq e \leq 1$, in good agreement 
with the numerical results
\cite{bmtjpa2001}.

The trace formula (\ref{dgscl}) also has the correct harmonic-oscillator (HO) 
limit for $E \rightarrow 0$, where Tr$\mathcal{M}_{\rm PO} \rightarrow 2$ and 
all coefficients in the expansion of the action phase in $Q$ and $p$ go to zero 
(and $\int d Q d p \rightarrow 2 \pi E$). The Poincar\'e variables 
$Q$ and $p$ become cyclic in this HO limit. In the spirit of the uniform 
approximations (see Sec.\ 6.3 of \cite{book} and App.\ A of \cite{bracktanaka}) 
within the ISPM, we may use a canonical transformation from the variables 
($Q,p$) to new variables ($\widetilde{Q},\widetilde{p}$) in which one has a simple 
analytical expression for the PO amplitude 
$\mathcal{A}_{\rm PO}^{G}\left[1-\exp(-E/A_{\rm PO}^{G})\right]$, instead of 
$A_{\rm PO}^{G}$ in (\ref{gutz}), with the two correct limits to the HO trace formula 
\cite{book} for $E\to 0$ and to the Gutzwiller trace formula (\ref{gutz})
for large $E$. We should note that this procedure is not unique. Strictly 
mathematically, the reduction of the full ${\widehat S}_{\rm ct}(Q,p,\vareps)$ 
(2) to any desired normal form by means 
of canonical transformations is generally not possible because, according 
to Arnold \cite{arnold}, ``formal series for canonical transformations reducing 
a system to normal form generally {\it diverge}''. On the other hand, within 
the ISPM, we use as ``normal forms'' equation (\ref{genfunact})  
for the generating function with expansion (\ref{genfunexp}) near the 
stationary points rather than near the 
bifurcations. A similarity to the normal form theory is manifested
if we put formally $Q^*=0$, $p^*=0$ in (\ref{genfunexp}) in the system
of coordinates related to the bifucation point reducing the non-local ISPM 
to its local approximation valid nearly the bifurcation points. 
Moreover, from a  more
pragmatic point of view, the details of the required canonical transformation 
do not matter for the SPM approximation in narrow regions of phase space around 
the critical points, i.e., in the limit $\hbar \rightarrow 0$ which in practice
corresponds to large particle numbers $N$ through the Fermi energy $\vareps_F$
at a rather small parameter $\alpha$ and larger 
averaging width $\gamma$ of the gross shell structure. 
We emphasize also the chaos-to-order transition of the PSS in the limit to the 
symmetry breaking point $e \rightarrow 0$ (see also Fig.\ \ref{fig1}).
In this limit, the isolated trajectories are transformed into the degenerate families 
of the periodic orbits. 

Expressions found from (\ref{dgscl}) locally for the separate bifurcations
of the $R$ or $L$ orbits are in agreement with the results
\cite{sie97,ssun97,ssun98,schomerus} obtained using the standard normal forms for
the pitchfork bifurcations. However, for the full cascade of bifurcations near 
the saddle energy of the HH potential, our result (\ref{dgscl}) goes beyond the 
normal-form theory. It is a continuous function through all bifurcation points
near the saddle energy and also down to the limit to the 
symmetry-breaking point 
at $E=0$. The coefficients $\varepsilon(E)$ and $a(E)$ in  (\ref{amphh}) are 
also continuous functions of the energy $E$ through all stationary points (POs).
Note also that our ISPM expression (\ref{dgscl}) for the shell correction to
the level density is a sum of separate contributions of all involved POs,
and a coarse-graining over the energy $\vareps$ (cf.\ below) may therefore be 
performed analytically. Thus, one has 
a possibility to study analytically both gross and fine shell structures.
This is in contrast to the results \cite{kaidelbrack,bracktanaka} using uniform 
approximations based on the normal-form 
theory \cite{sie97,ssun97,ssun98,schomerus}, where at each critical point all
involved POs give one common contribution. 

\section{ Discussion of results}

\begin{figure} 
\begin{center}\hspace*{-0.6cm}
\includegraphics[width=0.45\textwidth,clip]{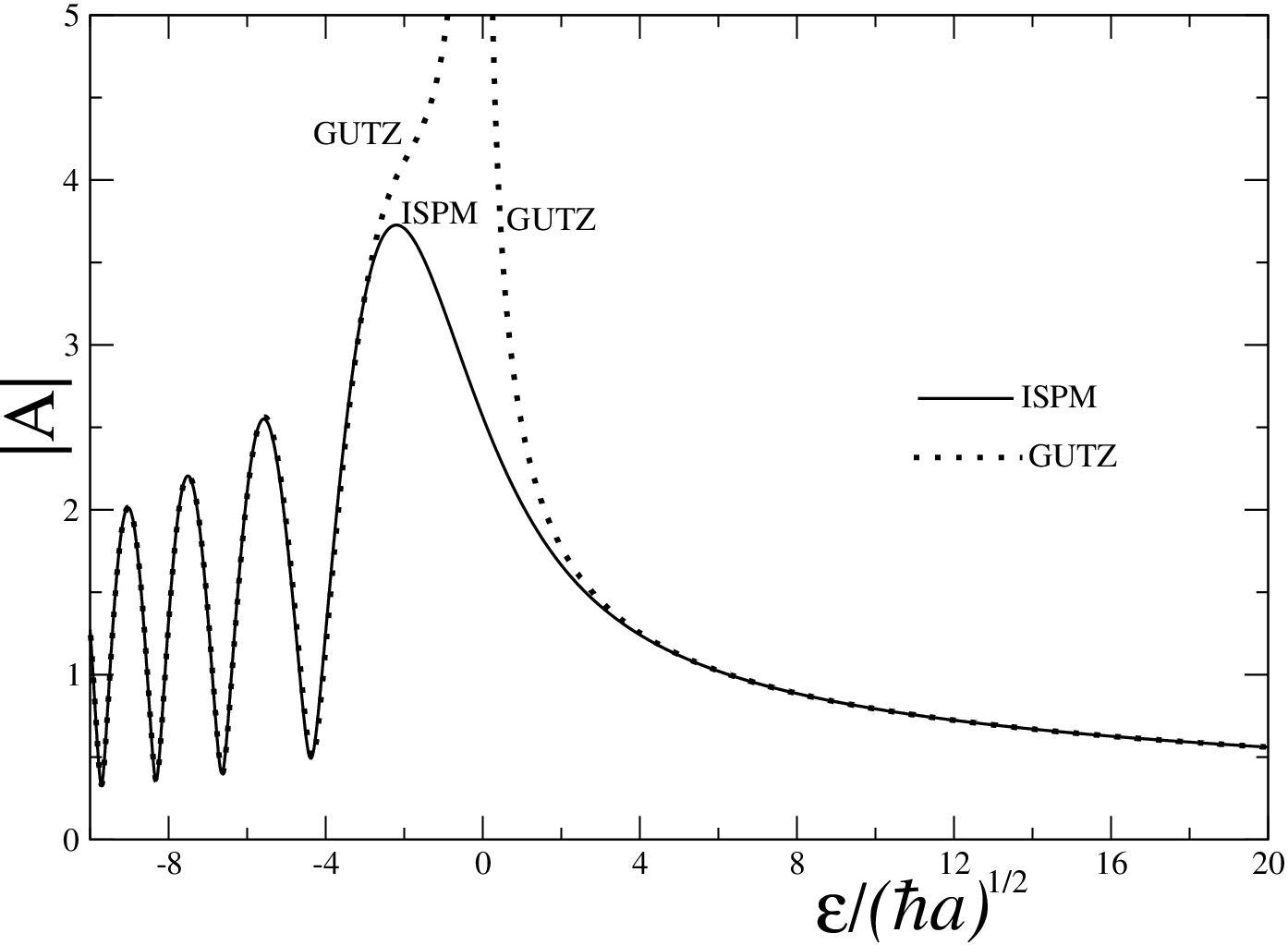}
\end{center}

\vspace{-0.2cm}
\caption{ The modulus of the 
amplitude factor $|\mathcal{A}|$ (\ref{amphh}) as a function of 
$\varepsilon/(\hbar a)^{1/2}$ is plotted around the bifurcation point. 
The convergence 
of the ISPM to the Gutzwiller amplitude (GUTZ) far from the bifurcation 
is clearly seen.
}
\label{fig3}
\end{figure}

Fig.\ \ref{fig3} shows the modulus of the 
amplitude factor $|\mathcal{A}|$ 
(\ref{amphh}) as function of 
$\xi=\varepsilon/(\hbar a)^{1/2}$. As seen from this figure,  there is a
typical enhancement of the amplitude $|\mathcal{A}|$ near the bifurcation
point $\xi=0$. The ISPM amplitude factor (\ref{amphh}) has a finite value
at the bifurcation point $\varepsilon=0$ and converges to the
Gutzwiller SSPM asymptotics ($\xi \gg 1$) rather rapidly for $\xi>0$.
As usual, one can see the characteristic ``ghost'' oscillations 
at negative $\xi$
which do not contribute into the semiclassical level density. 

For the purpose of studying the improved level density around
the bifurcation points, we consider a slightly averaged 
level density, thus avoiding the convergence problems that usually arise
when one is interested in a full semiclassical quantization.
Such a ``coarse-graining'' can be done by folding the level density over
a Gaussian of width $\gamma$ \cite{book,migdal}.
(The particular choice of a Gaussian form of the averaging 
function is immaterial
and guided only by mathematical simplicity.) 
Applying this procedure to the semiclassical
level density (\ref{dgscl}), one gets \cite{strutmag,book,migdal}
\begin{equation}
\delta g_{\gamma,sc}(E)
= \sum_{\rm PO} \delta g_{\rm sc}^{({\rm PO})}(E)\,
e^{-\left(\frac{\gamma T_{\rm PO}}{\hbar}\right)^2}.
\label{dgsclgamma}
\end{equation}

\begin{figure*}[tbp]
\begin{center}
\includegraphics[width=0.8\textwidth,clip]{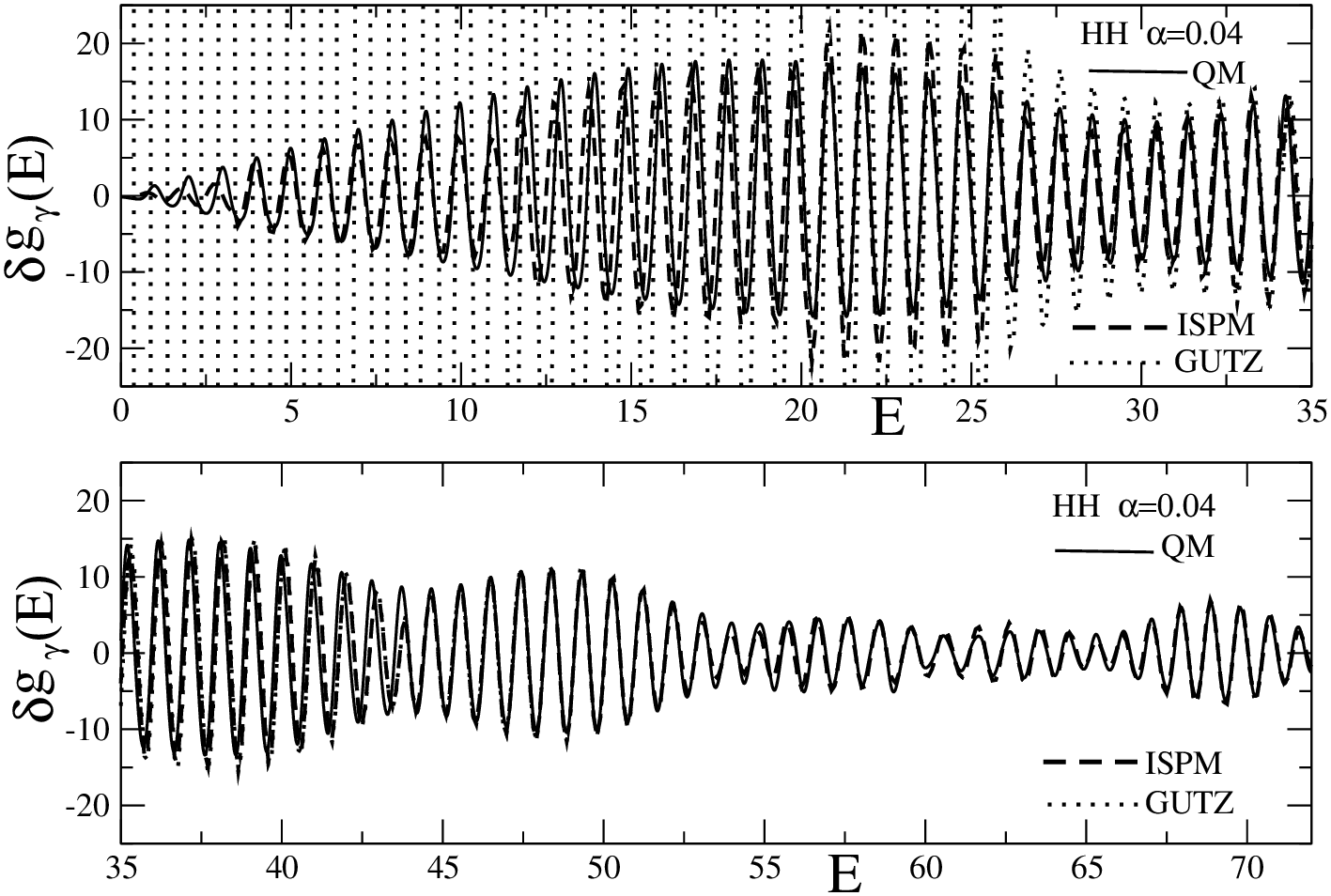}
\end{center}
\caption{ Quantum (QM, solid), semiclassical (ISPM, dashed), and
Gutzwiller (GUTZ, dots) level density versus energy
$E$ (in units of $\hbar\omega$). Only the primitive POs A, B and C
are included in the semiclassical calculations,
the Gaussian averaging width is $\gamma=0.25 \hbar \omega$.
}
\label{fig4}

\begin{center}
\includegraphics[width=0.8\textwidth,clip]{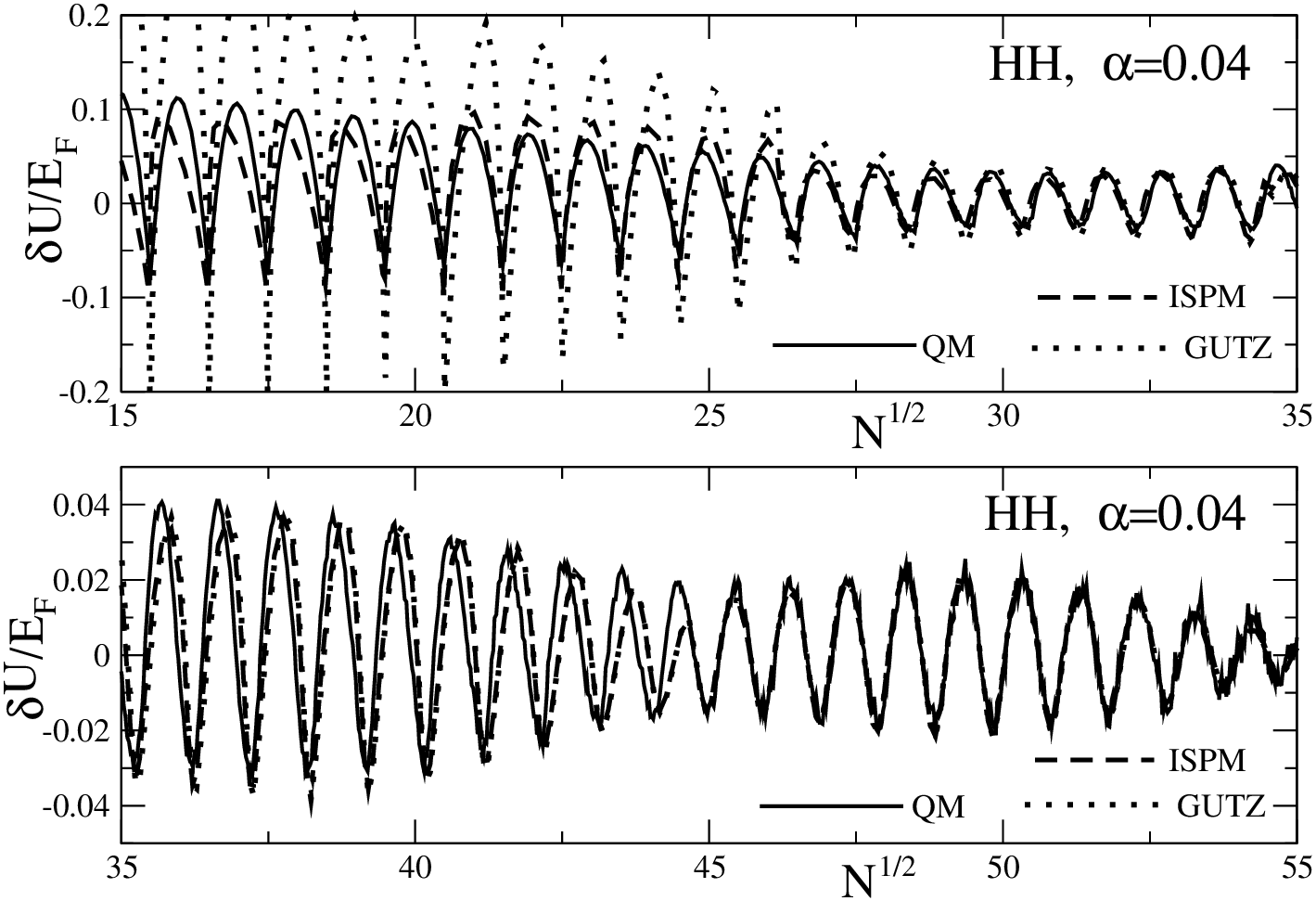}
\end{center}
\caption{ Quantum and semiclassical energy shell corrections $\delta U$ (\ref{descl1})
(in units of the Fermi energy $E_F$) versus particle number parameter $N^{1/2}$, with
$N = 2\int_0^{E_F} d E  \,g(E)$ (notation as in Fig.\ \ref{fig4}). 
The same primitive POs as in Fig. \ref{fig4} are included.
}
\label{fig5}
\end{figure*}

The averaging of the oscillating level density yields an exponential decrease
of the amplitudes with increasing periods $T_{\rm PO}$ and/or $\gamma $.  As
shown in \cite{mfaptp2006}, for $\gamma$ about 1/3 
(in $\hbar \omega$ units), all
large-action paths are strongly damped and only the time-shortest POs
contribute to the oscillating part of the level density, yielding its
gross-shell structure.  For a study of the bifurcation phenomenon,
however, we need smaller values of $\gamma$. In Figs.\ \ref{fig4} and \ref{fig5}
we used the coarse-grained Gutzwiller trace formula 
(\ref{gutz},\ref{dgsclgamma}) including the simplest primitive orbits 
$A, B=L_4$ and $C=R_3$. 

It is interesting that the gross-shell structure manifests itself
for the HH parameter $\alpha=0.04$ even for a relatively small averaging parameter 
$\gamma=0.25 \hbar \omega$. Therefore, we should expect also a good agreement
between semiclassical and quantum results for the energy shell corrections
$\delta U$ as functions of the particle numbers $N^{1/2}$ for the same $\alpha=0.04$ 
for larger energies (but still far enough from the bifurcation 
points, cf. Fig.\ \ref{fig5}).

The shell-correction energy $\delta U$,
i.e., the oscillating part of the total energy $U$ of a system of $N$
fermions occupying the lowest quantum levels in a given potential,
can be expressed in terms of the oscillating components 
$\delta g_{\rm sc}^{\rm PO}(E)$ at the Fermi energy 
$\vareps=\vareps_F$ of the semiclassical level density,
\be\label{gesclsum}
\delta g_{sc}(E)=\sum_{PO} \delta g_{sc}^{PO}(E)\,,
\ee
 as~\cite{strutmag,strusem,book,migdal}
\be\label{descl1}
\!\delta U = 2\sum_{\rm PO} \left(\frac{\hbar}{T_{\rm PO}}\right)^2
\delta g_{\rm sc}^{\rm PO}(E_F)\,.
\ee
Here, $T_{\rm PO}$ is the time of the motion along the 
PO including its repetitions at $\vareps=\vareps_F$, 
$T_{\rm PO} = M_{\rm PO} T_{\rm PPO}$ ($M_{\rm PO}$ is the repetition number).
We are taking into account the spin degeneracy factor 2 in (\ref{descl1}).
The semiclassical representation of the shell-correction energy
(\ref{descl1}) differs from that of $\delta g_{sc}(E)$ (\ref{gesclsum}) 
(at $E=E_F$) only by a factor
$(\hbar/T_{\rm PO})^2$ under the sum, which suppresses
contributions from orbits of larger time periods (actions). Thus the 
periodic orbits with smaller periods play a dominant role in determining 
the shell-correction energy \cite{strutmag,strusem}.
Finally, we should note that the higher the degeneracy of an orbit,
the larger the volume occupied by the orbit family in the phase space,
and also the smaller its time period (action), the more important is its
contribution to the level density.

Figs.\ \ref{fig4} and \ref{fig5} show a rather good agreement 
between the semiclassical and quantum results, in spite of 
using only the three shortest orbits $A$, $L_4$ ($B$), and $R_3$ (C). 
These are seen to yield the correct gross-shell 
structure for the parameter $\alpha=0.04$ and widths for the 
Gaussian averaging $\gamma=0.25\hbar\omega$ (or, similarly, for 
$\gamma=0.6\hbar\omega$) in the energy region below the saddle ($E=E_{barr}$) 
and above the bottom ($E=0$).
The discrepancies at smaller energies are related to the symmetry breaking 
at $E=0$, as discussed above, and will be removed when using our full ISPM 
trace formula (\ref{dgscl}). In the quantum-mechanical determination 
of $\delta U$ (see Ch.\ 4.7 of \cite{book} and \cite{migdal,mfaptp2006}
for discussions of the Strutinsky averaging method), the plateau condition 
for the averaged energy was satisfied for a Gaussian width $\tilde{\gamma} \simeq 
1.75 \hbar \omega$ and a curvature correction parameter $M=6$. 

\section{Conclusions}

We have obtained an analytical trace formula for the oscillating part 
of the level
density in H\'enon-Helies Hamiltonian as a sum over periodic orbits. 
It is continuous through all critical points, in particular here the 
harmonic oscillator 
limit at zero energy and the cascade of pitchfork bifurcations near 
the saddle energy.
We find an enhancement of the semiclassical amplitudes near the most 
critical points. 
The numerical agreement with quantum results is rather good,
in spite of a rather simple uniform ISPM approximation including only
the simplest primitive periodic orbits. The quantum-classical correspondence
for the chaos-order transitions is shown through the Poincar\'e surface of
sections in the limit from the non-integrable region of the energies to
the symmetry breaking point.
 Our semiclassical analysis may therefore lead 
to a deeper understanding of the shell structure effects in 
finite fermionic systems such as nuclei, metal clusters or 
semiconductor quantum dots whose conductance and magnetic 
susceptibilities are significantly modified by shell effects 
(see \cite{book,migdal,magvlasar2013,fkmsprb1998,belyaev} for
examples). 

\begin{ack}

We are grateful to K. Tanaka for
many discussions and help with the numerical calculations.
Some of the classical orbits and their properties were calculated 
using the program {\tt pos.for} developed by Ch. Amann in \cite{cam}.
We also thank J.P. Blocki, S.N. Fedotkin and A.P. Kobushkin   
for many useful discussions.
A.G.M.\ acknowledges the universities of Regensburg, Kyoto and Osaka (RCNP), 
the Nagoya Institute of Technology, the Japanese Society of Promotion of 
Sciences, and the National Centre of the Nuclear Research of Warsaw 
for very kind hospitality and financial support.  
\end{ack}

\appendix \setcounter{equation}{0}
\renewcommand{\theequation}{A\arabic{equation}}

\begin{center}
\textbf{Appendix A: The trace of the stability matrix for the A orbit}
\label{appA}
\end{center}

For small energies $e$, the trace $Tr\mathcal{M}_A$
can be expressed through Mathieu functions by using a general method
of soving Hill's equation for the Poincar\'e coordinate  $x(t)$ perpendicular 
to the A orbit directed along the $y$ axis (Fig.\ \ref{fig2}).
The perturbation $x(t)$ (in scaled variables (\ref{scaling})) near the 
A orbit is determined by Hill's equation
(\ref{eqmot}) for the HH Hamiltonian (\ref{scaling}),
\be\label{hilleq}
\ddot{x}(t) +\left[1+2 y_{A}(t)\right] x(t)=0,
\ee
where $ y_{A}(t)$ is the periodic solution for the A orbit 
\cite{bmtjpa2001,kaidelbrack,fmbpre2008},
\be\label{ya}
y_A=y_1 + (y_2-y_1) {\rm sn}^2(z,k),\quad z=a_kt+F(\varphi,k),
\ee
${\rm sn}(z,k)$ is the Jacobi elliptic function \cite{byrdbook}
with argument $z$; its modulus $k$ and the constant $a_k$ are given by
\be\label{ka}
k=\sqrt{\frac{y_2-y_1}{y_3-y_1}}, \qquad a_k=\sqrt{\frac{y_3-y_1}{6}}; 
\ee
$y_1$ and $y_2$
are the lower and upper turning points,
\bea\label{yn}
y_1&=&\frac12 - \cos\left(\frac{\pi}{3}-\frac{\phi}{3}\right),\qquad 
y_2=\frac12 - \cos\left(\frac{\pi}{3}+\frac{\phi}{3}\right),\nonumber\\
y_3&=&\frac12 + \cos\left(\frac{\phi}{3}\right), \qquad\qquad \cos\phi =1-2 e.
\eea
 $F(\varphi,k)$ is the incomplete elliptic integral of first
kind as a function of $\varphi=\arcsin \{[(y_0-y_1)/(y_2-y_1)]^{1/2}\}$,
$y_0=y_A(t=0)$ is the initial value. Using the Fourier expansion of ${\rm sn}^2(z,k)$, 
one has \cite{milne}
\begin{align}\label{foursn2}
{\rm sn}^2(z,k)=&
\frac{K(k)-E(k)}{k^2 K(k)} -\frac{2 \pi^2}{k^2 K^2(k)}\nonumber\\
&\times\sum_{n=1}^{\infty} 
\frac{n s^n}{1-s^{2 n}}\;\cos\left(\frac{\pi n z}{K(k)}\right),
\end{align}
where $s=\exp\left[-\pi K(\sqrt{1-k^2})/K(k)\right]$ is 
Jacobi's Nome \cite{byrdbook},
$K(k)$ and $E(k)$ are the complete elliptic integrals of first and second
kind, respectively. 

For small energies $e$ where Jacobi's Nome $s$ is small, 
$s \rightarrow k^2/16 \approx \sqrt{e}/(12 \sqrt{3})$ 
for $e \rightarrow 0$ ($k \rightarrow 0$, see (\ref{ka})), the convergence of
the Fourier series (\ref{foursn2}) is fast even for $e \siml 0.8$
($s\siml 0.04$).
For such energies, we may truncate the Fourier series approximately, keeping
only the first ($n=1$) harmonic term. After substitution of (\ref{ya})
with the expansion (\ref{foursn2}), a simple transformation of the time  
variable and the parameters in (\ref{hilleq}) leads to the standard equation 
for the Mathieu equation \cite{abramov}:
\be
\frac{{\rm d}^2}{{\rm d} \tau^{2}}x(\tau) +
\left[a -2 b \cos\left(2 \tau\right)\right]x(\tau) =0\,,
\ee
with 
\begin{align}\label{parammathieu}
\tau&= \pi z/[2 K(k)]\,,\nonumber\\
a&=\left(\frac{2 K}{\pi a_k}\right)^2
\left\{1+2\left[y_1 + (y_2-y_1)\frac{K-E}{k^2 K}\right]\right\},
\nonumber\\
b&= 8 s(y_2-y_1)/[k^2 a_k (1-s^2)]\,.
\end{align}
The solution of this second-order ordinary differential equation can be 
sought as a linear superposition of the fundamental set of the even  
$M_{C}(a,b,\tau)$ and odd $M_{S}(a,b,\tau)$  
Mathieu functions with arbitrary constants $C_1$ and 
$C_2$:
\be\label{mathsol}
x(\tau)= C_1 {\rm M}_{C}(a,b,\tau) + 
C_2  {\rm M}_{S}(a,b,\tau)\,.
\ee
Applying to (\ref{mathsol}) the boundary conditions for calculations 
of the stability matrix elements $\mathcal{M}_{xx}$ and  
$\mathcal{M}_{\dot{x}\dot{x}}$  as in \cite{fmbpre2008}, one 
obtains the constants $C_1$ and $C_2$ and the following diagonal 
matrix elements:
\begin{align}\label{matrixelem}
\mathcal{M}_{xx}&=
\frac{x(T)}{x(0)}\Big|_{\dot{x}(0) \rightarrow 0}=
\frac{{\rm M}_{S,0}'{\rm M}_{C,T} - {\rm M}_{C,0}'{\rm M}_{S,T}}{
{\rm M}_{C,0}{\rm M}_{S,0}' - {\rm M}_{S,0}{\rm M}_{C,0}'}\,,\nonumber\\
{}\hspace{-4.0ex}\mathcal{M}_{\dot{x}\dot{x}}&=
\frac{\dot{x}(T)}{\dot{x}(0)}\Big|_{x(0) \rightarrow 0}=
\frac{{\rm M}_{S,T}'{\rm M}_{C,0} - {\rm M}_{C,T}'{\rm M}_{S,0}}{
{\rm M}_{C,0}{\rm M}_{S,0}' - {\rm M}_{S,0}{\rm M}_{C,0}'}\,,
\end{align}
where primes means the partial derivatives of the Mathieu
functions $M_{C}(a,b,\tau)$ and $M_{S}(a,b,\tau)$ 
with respect to $\tau$. The lower indices $0$ and $T$ show the values
at the initial $t=0$ and final $t=T$ times, and $T=T_A=2 K(k)/a_k$ is the 
period of motion of the particle along the A orbit.
For the trace $Tr\mathcal{M}_A$, one finally finds 
\be\label{trma}
Tr\mathcal{M}_A=\mathcal{M}_{xx}+\mathcal{M}_{\dot{x}\dot{x}}\,,
\ee
with the diagonal matrix elements given in (\ref{matrixelem}).

For comparison, we recall the solution for the trace $Tr\mathcal{M}_A$ 
near the saddle $e \rightarrow 1$ obtained in \cite{bkwf2006,fmbpre2008} 
in terms of the Legendre functions by using in (\ref{ya}) the approximation
of  the Jacobi elliptic function, ${\rm sn}(z,k) \approx \tanh(z)$, i.e.,
by the zero-order term of its expansion near the saddle in powers of $1-k^2$
\cite{abramov}:
\bea\label{snexpsad}
{\rm sn}(z,k) \approx \tanh z \!\left[\!1 
+ 
\frac14 (1-k^2)\!\left(1 - \frac{z}{\sinh z \cosh z}\right)\!\right]\!\!.\;
\eea
As shown in \cite{bkwf2006,fmbpre2008}, the trace $Tr\mathcal{M}_A$ 
is in this approximation in good agreement with the numerical results 
\cite{bmtjpa2001} near the saddle $e \rightarrow 1$.

Generally speaking, for a more general solution,  
it is difficult to take into account exactly the next term of the expansion 
(\ref{snexpsad}) to get a simple analytical result similar to that presented 
explicitly in \cite{fmbpre2008}. 
However, we may use the approximate constant $r$ for the square brackets
in (\ref{snexpsad}), which effectively takes into account the correction to
$\tanh z$,
\be\label{rconst} 
r \approx 1+ r_{corr}(1-k^2),\qquad r_{corr}=1/4\;.
\ee
Within this approximation, one has again the result in 
terms of the Legendre functions $P_{\nu}^{\mu}$ and $Q_{\nu}^{\mu}$
with complex indices $\nu$ and $\mu$ depending on the energy $e$
\be\label{numu}
\mu=i \sqrt{A+B}\,, \qquad \nu=(-1+i\sqrt{4 A -1})/2\,,
\ee
where $B$ is the same as in \cite{fmbpre2008} but $A$ contains the 
additional constant factor $r$:
\be\label{ab}
A=12\,r\,k^2\,, \qquad B=(1+2 y_1)/a_k^2\;,
\ee
corresponding at $r$=1 (or $r_{corr}=0$ in our notations)
to the results in \cite{fmbpre2008}.
\begin{figure}
\begin{center}
\includegraphics[width=0.46\textwidth,clip]{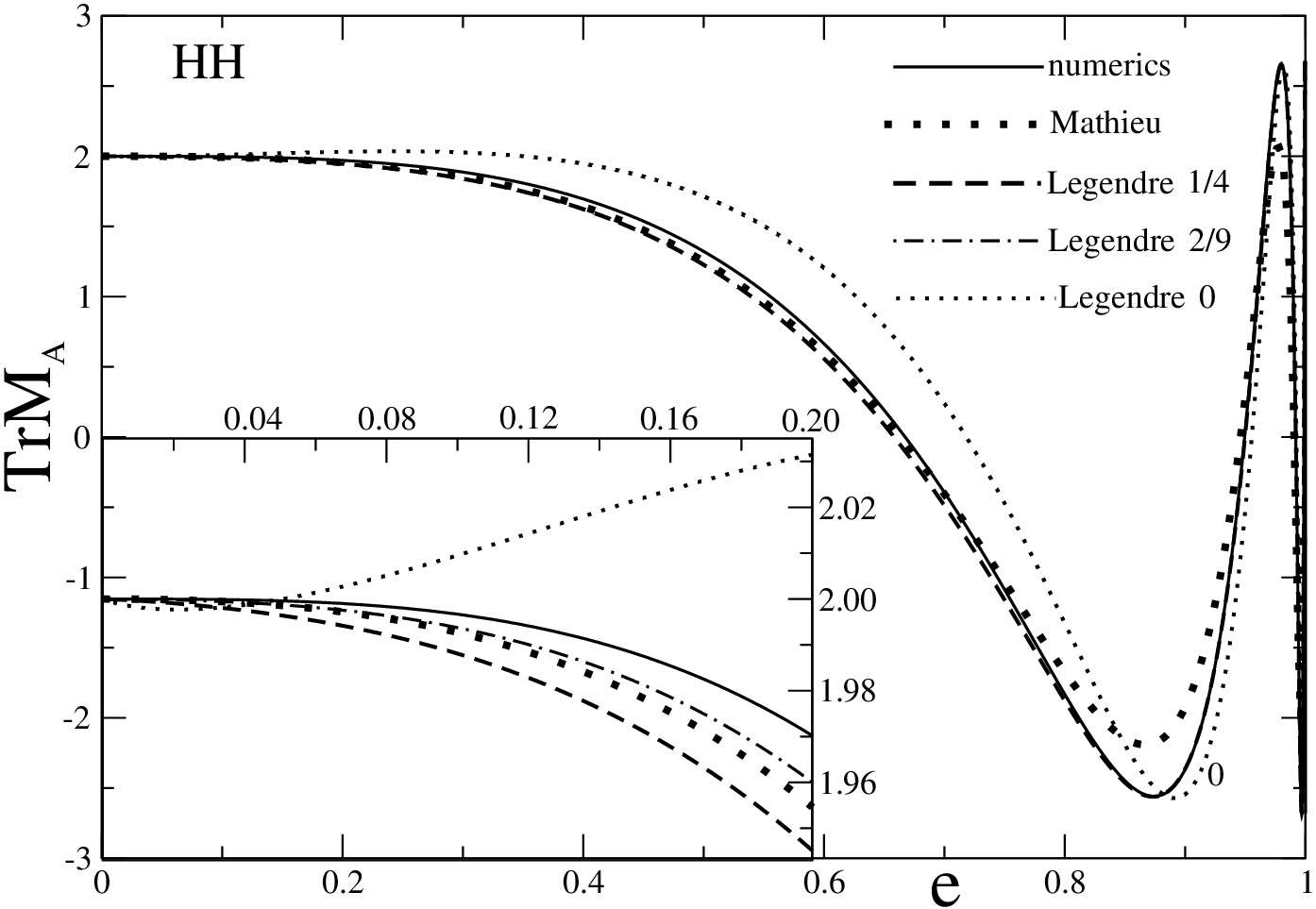}
\end{center}
\caption{Numerical and analytical traces $TrM_A$ for A orbit. 
The solid line is the 
full numerical result (after \cite{bmtjpa2001,bracktanaka}); 
heavy dots show the 
approximation (\ref{trma}) through the Mathieu functions
(\ref{matrixelem}); the dashed line shows the improved Legendre function 
approximation with the constant (\ref{rconst}). 
Light dots present the asymptotic Legendre function 
approximation with $r=1$ ($r_{corr}=0$). The insert shows
a small region near the symmetry breaking limit $e\to0$; 
the values of the constants $r_{corr}=0, 1/4$ and $2/9$ (\ref{rconst}) 
for different Legendre function approximations are shown.
}
\label{fig6}
\end{figure}
Fig.\ \ref{fig6} shows the comparison of numerical calculations 
\cite{bracktanaka,bmtjpa2001} with our analytical results for the trace of the 
stability matrix $Tr\mathcal{M}_A$ in the case of the A orbit. As seen from 
Fig.\ \ref{fig6}, the solution for  $Tr\mathcal{M}_A$ in terms of
the Mathieu functions is in good agreement with the exact numerical 
results even at energies $e \siml 0.8$. We show there also another approximation 
in terms of Legendre functions with the indices (\ref{numu}), 
improved at finite and small energies $e$ through the constant $A$ (\ref{ab}) 
with $r$ given in (\ref{rconst}) as compared to the result ($r=1$) obtained earlier
near the saddle [i.e., using only the leading term in the expansion 
(\ref{snexpsad}) for $e-1 \ll 1$] \cite{fmbpre2008}. Through a modification of 
only one constant $r$ (\ref{rconst}), one has a remarkable agreement between 
this improved Legendre approximation and the numerical results everywhere
from the saddle point $e$  to the harmonic oscillator limit 
$Tr\mathcal{M}_A \rightarrow 2$ for $e \rightarrow 0$ (Fig.\ \ref{fig6}).

As shown in the insert of Figure \ref{fig6},
this approximation can be slightly improved changing the constant $r_{corr}$
in (\ref{rconst}) from $r_{corr}=1/4$ ($z \rightarrow \infty$) to about 2/9
of finite values of $z$. For small energies $e$ ($k \rightarrow 0$), 
one can, again, formally use (\ref{snexpsad}): the correction to $\tanh z$ can 
be neglected for small times ($z \ll 1$) because it gives the dominating 
contribution to $TrM_A$ [equation (\ref{hilleq}) with (\ref{ya}) becomes
approximately the same at small $z$], and $TrM_A \rightarrow 2$ in all analytical 
approximations, in agreement with the numerical results. In the limit 
$e \rightarrow 0$ the Legendre function approximation converges, indeed, to the 
analytical Mathieu function solution (see the insert).
Note also that this agreement with the numerical results
is not sensitive to a variation of the constant 
$r_{corr}$ around the analytical value (\ref{rconst}). 
The particle moving near the 
A orbit spends much more time near the saddle where the function of $z \propto 
t$ in the circle brackets (\ref{snexpsad}) is almost constant with respect to 
the remaining part of the trajectory. 
However, at small energies $e \rightarrow 0$, one finds a smaller 
time region ($z \ll 1$) where the correction in (\ref{snexpsad}) becomes 
negligible for $TrM_A$, such that all curves in Fig.\ \ref{fig6} have the same correct 
harmonic oscillator limit $2$. Thus, a rather complicated function 
of time in the correction to the leading (hyperbolic tangent) term of the 
expansion of the Jacobi function (\ref{snexpsad}) can be reduced to a form 
involving the same Legendre functions as in \cite{fmbpre2008}, 
but with modified 
indices by the constant $r$ (\ref{rconst}) through (\ref{ab}).


\begin{thebibliography}{99}\itemsep=-4pt

\bibitem{gutz} Gutzwiller M 1990 {\it Chaos in Classical and Quantum
               Mechanics} (Springer-Verlag, New York)

\bibitem{strut} Strutinsky V M  1975 Nukleonika {\bf 20} 679                 

\bibitem{strutmag} Strutinsky V M and Magner A G 1976 Sov. Phys. Part.
                   \& Nucl. {\bf 7} 138

\bibitem{strusem} Strutinsky V M, Magner A G, Ofengenden S R and 
                  D{\o}ssing T 1977 Z. Phys. {\bf A 283} 269

\bibitem{book} Brack M and Bhaduri R K 2003 {\it Semiclassical Physics}; 
               Frontiers in Physics No 96, 2nd ed. (Westview Press, Boulder, CO)

\bibitem{migdal} Magner A G, Yatsyshyn I S,
                 Arita K and Brack M 2011 Phys. Atom. Nucl., {\bf 74} 1445

\bibitem{hh} H\'enon M and Heiles C 1964 Astr. J. {\bf 69} 73

\bibitem{kaidelbrack} J. Kaidel and M. Brack 2006 Phys. Rev. E {\bf 70} 016206

\bibitem{bracktanaka} M. Brack and K. Tanaka 2008 Phys. Rev. E {\bf 77} 046205 

\bibitem{sie97}   Sieber M 1997 J. Phys. A: Math. Gen. {\bf 30} 4563.

\bibitem{ssun97}  Schomerus H and Sieber M 1997 J. Phys. A: Math. Gen. {\bf 30} 4537

\bibitem{ssun98}  Sieber M and Schomerus H 1998 J. Phys. A: Math. Gen. {\bf 31} 165

\bibitem{schomerus} Schomerus H 1998 J. Phys. A: Math. Gen. {\bf 31} 4167

\bibitem{mfaptp2006} Magner A G, Fedotkin S N and Arita K 2006
                     Prog. Theor. Phys. {\bf 115} 523 

\bibitem{hhprl} Brack M, Bhaduri R K, Law J and Murthy M V N 1993 
                Phys. Rev. Lett. {\bf 70} 568

\bibitem{BMYnpae2010} Blocki J P, Magner A G and Yatsyshyn I S 2010 
                      At. Nucl. Energy {\bf 11} 239

\bibitem{BMYijmp2011} Blocki J P, Magner A G and Yatsyshyn I S 2011 
                      Int. J. Mod. Phys. E 20, 292 (2011)

\bibitem{BMprc2012} Blocki J P and Magner A G 2012 Phys. Rev. C {\bf 85} 064311 

\bibitem{cam} Amann Ch and Brack M 2002
              J. Phys. A: Math. Gen. {\bf 35} 6009

\bibitem{bmtjpa2001} Brack M, Mehta M and Tanaka K 2001 J. Phys. A:
                     Math. Gen. {\bf 34} 8199 

\bibitem{bkwf2006} Brack M, Kaidel J, Winkler P and Fedotkin S N 2006
                   Few-Body Syst. {\bf 38} 147

\bibitem{fmbpre2008} Fedotkin S N, Magner A G and Brack M  2008
                     Phys. Rev. E {\bf 77} 066219 

\bibitem{arnold}  Arnold V I 1989 {\it Mathematical Methods of Classical 
                  Mechanics} (New York: Springer-Verlag, 2nd edition),
                  Addition 7

\bibitem{magvlasar2013} Magner A G, Vlasenko A A and Arita K 2013 
                        Phys. Rev. E {\bf 87} 062916 

\bibitem{fkmsprb1998} Frauendorf S, Kolomietz V M, Magner A G
                      and Sanzhur A I 1998 Phys. Rev. B {\bf 58} 5622

\bibitem{belyaev} Magner A G, Gorpinchenko D V and Bartel J 2014
                  Phys. At. Nucl. {\bf 77} 1229 

\bibitem{byrdbook} Byrd P F and Friedman M D 1971 {\it Handbook of Elliptic 
                   Integrals for Engineers and Scientists} 
                   (2nd ed, Sprinder-Verlag New York Heidelberg Berlin)

\bibitem{milne} Milne S C 2002 {\it Infinite Families of Exact Sumes of Squares 
                Formulas Jacobi Elliptic Functions, Continued Fractions, 
                and Schur Functions} (Kluwer Academic Publishers); 
                the Ramanujan Journal {\bf 6} 7

\bibitem{abramov} Abramovitz M and Stegun I A 1965 {\it Handbook of 
                  mathematical functions} (Dover Publications, New York)

\end{thebibliography}
\end{document}